\documentstyle[aps,epsf,twocolumn]{revtex}

\begin{document}

\draft

\preprint{TPR-97-33}

\title{Bose-Einstein Correlations from Random Walk Models}
\author{Boris Tom\'a\v{s}ik$^{a,b}$, Ulrich Heinz$^a$, 
        and J\'an Pi\v{s}\'ut$^{b,c}$}
\address{$^a$Institut f\"ur Theoretische Physik, Universit\"at Regensburg, 
             D-93040 Regensburg, Germany\\
        $^b$Faculty of Mathematics and Physics, Comenius University, 
            Mlynsk\'a Dolina, SK-84215 Bratislava, Slovakia\\
        $^c$Laboratoire de Physique Corpusculaire, Universit\'e Blaise Pascal,
            Clermont-Ferrand, F-63177 Aubi\`ere, Cedex, France}
\date{January 19, 1997}

\maketitle

\begin{abstract}

We argue that strong final state rescattering among the secondary
particles created in relativistic heavy ion collisions is essential to
understand the measured Bose-Einstein correlations. The recently
suggested ``random walk models'' which contain only initial state
scattering are unable to reproduce the measured magnitude and
$K_\perp$-dependence of $R_\perp$ in Pb+Pb collisions and the increase
of $R_l$ with increasing size of the collision system. 

\end{abstract}


 

An important aspect of understanding ultrarelativistic heavy ion
collisions is the clear identification of genuine collective nuclear
effects which cannot be explained in terms of a simple superposition
of nucleon-nucleon collisions. An integral part of our task to look
for new physics must therefore be the careful construction of models
for nucleus-nucleus collisions (``$A$-$B$ collisions'') based on a
superposition of individual nucleon-nucleon ($N$-$N$) collisions in
order to establish where they fail. 

Recently there have been some renewed attempts to construct such
models. In the ``Random Walk Model'' (RWM) of \cite{LNS97}
the single-particle transverse mass spectra measured in $A$-$B$
collision have been calculated by extrapolating those from $pA$
collisions. The LEXUS 
model of \cite{JK97} goes even further, by simulating $A$-$B$
collisions as a simple folding of independent $N$-$N$ collisions.  

Here we demonstrate that, while these models have had some success in
describing measured single-particle spectra, they fail to reproduce
crucial features of the observed two-particle Bose-Einstein
correlations. This is shown to be due to their lack of final 
state rescattering.

The idea behind the RWM was to provide an alternative interpretation
of the measured single-particle $m_\perp$ spectra, opposite in spirit
to the popular hydrodynamical parametrizations \cite{LHS90}. The
latter are based on the assumption of a locally thermalized hadron
resonance gas which undergoes longitudinal and transverse 
hydrodynamic expansion. The transverse expansion is interpreted as a
genuine nuclear collective effect with no analogue in $N$-$N$
collisions; it is identified through the characteristic flattening it
causes in the transverse mass spectra, especially at small $M_\perp <
2 m_0$ where the inverse slope parameter is found to increase linearly
with the rest mass $m_0$ of the produced hadrons \cite{LHS90,NA44S}.  

The RWM, on the other hand, starts from the observation that such a
flattening, relative to $N$-$N$ collisions, happens even in
$pA$ collisions where hydrodynamic transverse expansion is not
expected to occur. In the RWM an incident nucleon undergoes multiple
collisions in the nuclear target, leading to a random walk pattern in
the transverse momentum plane. When it collides inelastically it
creates a little ``fireball'', identical to those formed in elementary
inelastic $pp$ collisions, but moving with a transverse rapidity
$\rho$ which is Gaussian distributed. In $A$-$B$ collisions the same
mechanism works for both projectile and target nucleons. The width of
the $\rho$ distribution is fixed in $pA$ collisions and then
extrapolated to $A$-$B$ collisions using geometric considerations
\cite{LNS97}. Resonance decays are neglected, limiting the
applicability of the model to sufficiently large transverse momenta.  

In LEXUS \cite{JK97} the same idea is implemented more directly in
terms of a folding algorithm for $N$-$N$ collisions which, in contrast
to RWM, also includes longitudinal momentum degradation in multiple
collisions. However, the latter is not coupled to the broadening of
the $m_\perp$-distributions which is parametrized in terms of an
effective temperature (inverse slope) which again follows a simple
random walk rule.    

In both models the transverse broadening parameter is taken as
independent of rapidity and not correlated with the position of the
collision point (fireball) within the reaction zone. Secondaries do
not interact, and (except for energy degradation effects) all primary
collisions are alike. For this reason neither model can explain the
different chemical composition in $pp$ and $A$-$B$
collisions (see \cite{JK97}); they both make statements only about the
shape of the momentum distributions. 

Existing tests of the models are still very superficial. In
\cite{JK97} the LEXUS output was compared to pion and proton rapidity 
spectra and midrapidity transverse mass spectra from 200 A GeV S+S
collisions at the SPS. In Refs.~\cite{EHX97,ACRS97} recent data on
single-particle $m_\perp$-spectra from 160 A GeV Pb+Pb collisions at
the SPS, taken by the NA44 \cite{NA44S} and NA49 \cite{NA49S}
collaborations, were compared to the RWM and to hydrodynamical
parametrizations. Both types of models give a reasonable description
of the data; discrepancies occur mostly at low $p_\perp$, where
resonance decays distort the pion spectra and hydrodynamical flow
effects show up most strongly for the heavier hadrons. In particular,
the RWM has difficulties \cite{EHX97} to reproduce the strong rest
mass dependence of the transverse slope parameters published by NA44
\cite{NA44S} which have been quoted as evidence for collective
transverse flow \cite{LHS90,NA44S}. It is clear from these papers,
however, that without a systematic study of the $m_\perp$-spectra as a
function of the collision system and of rapidity, carefully including
resonance decays, single particle spectra alone may not lead to
definite conclusions about the validity of the RWM approach. In the
present note we therefore examine the discriminating power of
two-particle Bose-Einstein correlations. 

For ease of language our discussion will be based on the RWM, with
comments on LEXUS to follow below. To calculate the correlation
function for pairs of identical particles we need to know the emission
function $S_{\rm rw}(x,p)$ describing the phase-space distribution of
the particles emitted from the source. Since both the RWM and LEXUS
provide explicitly only momentum space information, we must try to
reconstruct the corresponding coordinate space information from the
description of the models given in Refs.~\cite{LNS97,JK97}. From the 
emission function the correlator is calculated via \cite{S73,P84,CH94} 
\begin{equation}
\label{1a}
        C_{\rm rw}(\bbox{q},\bbox{K}) = 1 +
        \frac {{\left | \int d^4x \, S_{\rm rw}(x,K)
        \, e^{iq\cdot x} \right |}^2}
        {{\left | \int d^4x \, S_{\rm rw}(x,K) \, \right | }^2 } \, ,
\end{equation}
where $K=(p_1+p_2)/2$ is the average pair momentum and $q = p_1-p_2$
the difference between the two observed on-shell momenta. Since $p_1$
and $p_2$ are on-shell, $q$ and $K$ satisfy the constraint 
 \begin{equation}
 \label{1b}
   q\cdot K = 0 \quad \Longrightarrow \quad q^0 
            = \bbox{q}\cdot {\bbox{K} \over K^0} 
            = \bbox{q}\cdot {\bbox{p}_1 + \bbox{p}_2 \over E_1 + E_2}\, .
 \end{equation}
The single-particle spectrum is calculated as
 \begin{equation}
 \label{1}
        E_p \frac{dN}{d^3p} = P_1(\bbox{p})
        = \int  d^4x \, S_{\rm rw}(x,\bbox{p},E_p) \, .
 \end{equation}
For the RWM it was given in the form \cite{LNS97}
 \begin{eqnarray}
        2\frac{dN}{dy\, dp_\perp^2 \, d\phi}
        &=& \int dY \, d\rho \, d\Phi \, f(\rho) \, \theta(Y_L - |Y-Y_0|)
 \nonumber \\ 
        &&\times \, \frac{V_0 \, m_\perp}{(2\pi)^3}\,\cosh (Y-y)\,
        \exp\left( - \frac{p\cdot u}{T}\right) \, \, .
 \label{2}
 \end{eqnarray}
Here $y,p_\perp,\phi$ describe the momentum $\bbox{p}$ of the
mea\-sured particle, $T$ is the temperature of the fireballs formed in
the individual $N$-$N$ collisions, and $Y$ is their longitudinal
rapidity which is distributed with a box distribution
$\theta(Y_L{-}|Y{-}Y_0|)$ between $Y_0{-}Y_L$ and $Y_0{+}Y_L$. $u$ is
the 4-velocity of the fireball, parametrized according to
$u(Y,\rho,\Phi) = (\cosh\rho\cosh Y$, $\sinh\rho \cos\Phi$, 
$\sinh\rho\,\sin\Phi$, $\cosh\rho\sinh Y) = \gamma(1,\bbox{u})$ where
$\rho$ is the transverse rapidity of the fireball and $\Phi$ its
direction. This gives 
 \begin{equation}
 \label{bterm}
        p \cdot u{=}m_\perp \cosh (Y{-}y)\cosh \rho - p_\perp
        \cos (\Phi{-}\phi) \sinh\rho \, .
 \end{equation}
The transverse fireball rapidity $\rho$ is distributed with
 \begin{equation}
 \label{3}
        f(\rho) = f_{AB}(\rho) =  \frac{2}{\sqrt{\pi \delta^2_{AB}}} \,
        \exp\left(
        - \frac{\rho^2}{\delta^2_{AB}} \right)\, .
 \end{equation}
The width of this distribution depends on the collision system $A$+$B$
via the random walk rule
 \begin{equation}
 \label{4}
        \delta^2_{AB} = (N_A + N_B - 2) \, \delta^2 \, ,
 \end{equation}
where $N_A$ ($N_B$) is the {\em average} number of nucleons in nucleus
$A$ ($B$) hit by a nucleon from nucleus $B$ ($A$). If the projectile
is just one nucleon, $N_B = 1$. The parameter $\delta$ is obtained by
fitting $pA$ data ($N_B=1$).

If one tries to reconstruct $S_{\rm rw}$ by comparing (\ref{1}) with
(\ref{2}) one is faced by certain ambiguities, although the appearance 
of a simple volume factor $V_0$ in front of the integral over the
fireball 4-velocity $u$ in (\ref{2}) excludes most functions 
$S_{\rm rw}(x,p)$ with complicated $x$-dependences. The simplest
assumption that the longitudinal and transverse fireball rapidities
$Y$ and $\rho$ of the fireballs are not correlated with the fireball
positions is, however, untenable. In this case the emission 
function factorizes, $S_{\rm rw}(x,p)=F(x) \cdot I(p)$ (with $I(p)$
given by the r.h.s. of Eq.~(\ref{2}) divided by $V_0$), and (\ref{1a})
yields a correlator which does not depend on $K$. Since all
heavy-ion data show a clear dependence of the correlation function
(in particular of its longitudinal width parameter $R_l$) on
the pair momentum $K$ \cite{NA49C,NA35C,NA44C}, such an emission
function is excluded. 

More realistically one should at least implement the simple
expectation that (in RWM language) fast fireballs live longer and fly
farther before decaying than slow ones. This leads to a correlation
between the decaying fireball's position $x$ and velocity $u$, inducing
also a correlation between $x$ and $p$ via the Boltzmann factor and
thereby a $K$-dependence of the correlator. We thus write
 \begin{eqnarray}
        S_{\rm rw}(x,p)
        &=& \int dY \, d\rho \, d\Phi \, F(x;Y,\rho) 
 \nonumber \\ 
        &&\times\, \frac{m_\perp}{(2\pi)^3}\,\cosh (Y-y)\,
        \exp \left( - \frac{p\cdot u}{T}\right)\, ,
 \label{7}
 \end{eqnarray}
with $\int d^4x\, F(x;Y,\rho) = V_0\, \theta(Y_L{-}|Y{-}Y_0|)$. 
Azimuthal symmetry excludes a $\Phi$-dependence of $F$. We will use
longitudinally boost-invariant space-time coordinates $x^\mu =
(\tau,\eta,r,\varphi)$, with $\tau^2=t^2-z^2$, $\eta = {1\over 2}
\ln[(t+z)/(t-z)]$, and integration measure $d^4x = \tau\, d\tau\,
d\eta\, r\, dr\, d\varphi$.  

In the RWM \cite{LNS97} the transverse fireball rapidity is not
correlated with its space-time position since the width $\delta_{AB}$
of $f(\rho)$ is determined from the {\em impact parameter averaged}
number of hit nucleons. Thus there are no {\em transverse} $x$-$p$
correlations in the source. This is easily seen to lead to a
correlator whose transverse width, given by the transverse HBT radius
$R_\perp$, does not depend on $K$. This contradicts the experimental
observation of a clear and rather strong decrease of $R_\perp$ as a
function of $K_\perp$ in 160 A GeV Pb+Pb collisions \cite{NA49C} which
cannot be explained by resonance decays \cite{H96,WH97}.  

One can try to include the missing transverse $x$-$p$ correlations by
making the broadening parameter $\delta_{AB}$ $r$-dependent. Instead
of using the impact parameter averaged number of hit nucleons from the
original formulation of the RWM \cite{LNS97,fn1},  
 \begin{equation}
 \label{c1}
   N_A = (2/3)\, \pi r_0^2 \, 2R_A \, n_0 \, ,
 \end{equation}
(where $r_0 = 0.8$ fm is the nucleon radius, $R_A = 1.12\, A^{1/3}$ fm
is the nuclear radius, and $n_0 = 0.17$ fm$^{-3}$ is the standard
nuclear density), we can take the actual value at distance $\bbox{r}$
from the collision axis:
 \begin{equation}
 \label{c2}
   N_A(\bbox{r}) = (\pi r_0^2)\, 2{\textstyle{\sqrt{R_A^2 - r^2}}}
   \, n_0 \, .
 \end{equation}
Via Eq.~(\ref{4}) this yields an $r$-dependent width $\delta_{AB}$
and, via (\ref{3}), an $r$-dependent $\rho$-distribution
$f_{AB}(\rho,r)$. We will now study whether the resulting transverse
$x$-$p$ correlations can cause the measured $K_\perp$-dependence of
$R_\perp$. 

Since the RWM does not take into account longitudinal energy
degradation by multiple initial state scattering, we can factorize
the $\rho$ and $Y$ dependence of the function $F$ in (\ref{7}):
$F(x;Y,\rho) = F'(x;Y)\,f(\rho,r)$. There is no correlation between
the transverse fireball rapidity and its longitudinal position. 
Furthermore, without longitudinal momentum loss the maximum and
minimum longitudinal fireball rapidities $Y_0\pm Y_L$ are not related
to the actual number of hit nucleons and thus independent of
$\bbox{r}$. Hence $F'(x;Y)$ can be further factored into
$H(\eta,\tau;Y)\cdot G(r,\varphi)$. The transverse HBT radius
$R_\perp$ can be calculated from~\cite{HTWW96} 
 \begin{equation}
 \label{c4}
   R_\perp^2 = \langle r^2 \sin^2 \varphi \rangle, \quad
   \langle f(x) \rangle = \frac{\int d^4x \, f(x) \, S(x,K)}
                                {\int d^4x \, S(x,K)}  .
 \end{equation}
Since the radial integral couples $G$ and $f$, we
must know the functional form of $G(r,\varphi)$. We take $G$
proportional to the number of collisions at distance $\bbox{r}$ from
the collision axis; for nuclear collisions at zero impact parameter this
is given by 
 \begin{equation}
 \label{c3}
   G(r,\varphi) = G(r) \propto \sqrt{R_A^2 - r^2} \sqrt{R_B^2 - r^2} \, .
 \end{equation}
In the following we consider only symmetric collisions, $R_A =R_B$. 
The generalization to the case $R_A \ne R_B$ is obvious. 

The evaluation of (\ref{c4}) is complicated by the fact that
Eq.~(\ref{bterm}) couples $\rho$ to $Y$ such that the function
$H(\eta,\tau;Y)$ must also be known. To obtain an as much as possible
model independent estimate for the behaviour of $R_\perp$ we use the
following simplification: We go to the LCMS (Longitudinally Co-Moving
System, $y = 0$) and {\em fix} $Y=0$, thus taking into account pion
production only from fireballs which rest in the LCMS. Since
contributions from other fireballs are suppressed by a factor
$\exp[-M_\perp (\cosh Y-1) \cosh\rho /T]$, we expect our
simplification to provide a good estimate for the correct
$R_\perp$. The advantage of this approximation is that now the
integration in Eq.~(\ref{c4}) factorizes in $(\tau,\eta)$ 
and $(r,\varphi)$. One finds
 \begin{equation}
 \label{c6}
   R^2_\perp \approx \frac{\int d\rho \int r\, dr \, r^2 \, C(r,\rho)}
                      {\int d\rho \int r\, dr \, 2 \,   C(r,\rho)}
 \end{equation}
where
 \begin{eqnarray}
   C(r,\rho)& = & (R_A^2 - r^2) 
   \left( \pi \left(4\pi\, r_0^2 \, n_0 \, 
                    {\textstyle{\sqrt{R_A^2 - r^2}}} - 2 \right) 
          \delta^2 
   \right )^{- \frac12}
 \nonumber \\
   && \times \exp\left[ - \frac{\rho^2} {(4\pi\, r_0^2 \, n_0 \, 
      \sqrt{R_A^2 - r^2} - 2 ) \delta^2} \right]
 \nonumber \\ 
   && \times \exp\left( -\frac{M_\perp}{T} \cosh \rho \right) \,
        \mbox{I}_0 \left( \frac{K_\perp}{T} \sinh \rho \right) \, .
 \label{c7}
 \end{eqnarray}
Here the integrations over $\Phi$ and $\varphi$ have already been
performed. Note that the $r$-integration in Eq.~(\ref{c6}) does not
extend to $R_A$ but only to $R_A - \varepsilon$ where $\varepsilon$ is
chosen such that at transverse distance $r = R_A - \varepsilon$ there
is still one nucleon-nucleon collision. This is ensured by requiring 
$\delta_{AA}(R_A - \varepsilon) = 0$; if $r > R_A - \varepsilon$ then
$\delta_{AA}(r)$ is negative due to Eqs.~(\ref{4}) and (\ref{c2}).
 \begin{figure}[h]\epsfxsize=8cm
 \centerline{\epsfbox{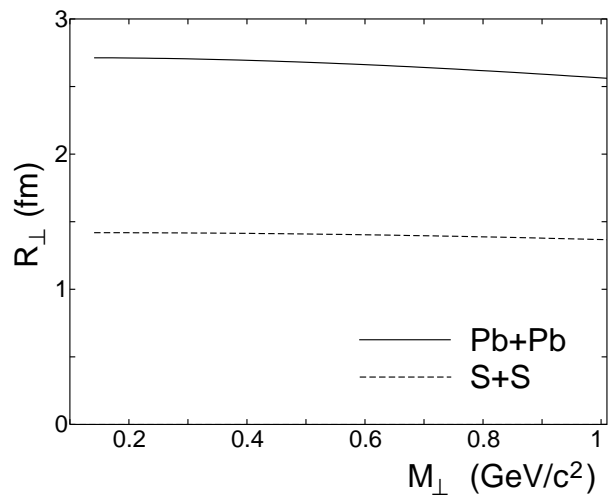}}
 \caption{The $M_\perp$-dependence of $R_\perp$ according to
   (\protect\ref{c6}) for Pb+Pb ($A$=207) and S+S ($A$=32) collisions.}
 \label{fig1}
 \end{figure}

For our computation we re-calculated $\delta$ according to
Eq.~(\ref{c1}) (see \cite{fn1}). Using $\delta_{pW} =0.3$ \cite{LNS97}
we got $\delta = 0.217$. Results for $R_\perp$ according to the
approximation (\ref{c6}) for S+S ($A$=32) and Pb+Pb ($A$=207)
reactions are shown in Fig.~\ref{fig1}. It is immediately clear that
the calculated values for $R_\perp$ {\em underpredict} the experimental
data (which, at small $M_\perp$, give $R_\perp \approx 6$ fm for Pb+Pb
\cite{NA49C} and $\approx 4$ fm for S+S \cite{NA35C,NA44C}) {\em
  considerably}. It has been argued in \cite{Uli-Rio} that the present
data on $R_\perp$ from Pb+Pb collisions {\em require} a strong transverse
expansion since the measured values of $R_\perp$ signal a source much
bigger than the original Pb nucleus. The $R_\perp$-values in
Fig.~\ref{fig1}, however, are consistent with a transversally
non-expanding source. Furthermore, the $M_\perp$-dependence of $R_\perp$
induced by the $r-$dependence of $\delta_{AB}$ is much too weak and
cannot reproduce the strong $M_\perp$-dependence seen in the Pb+Pb
data \cite{NA49C}. 

The wrong magnitude and $K$-independence of $R_\perp$ are, however, not
the only problems of the RWM/LEXUS models. Following \cite{PPZ97} we
now argue that they are also unable to correctly describe the
dependence of the longitudinal radius parameter $R_l$ on the size of
the collision system.   

In RWM and LEXUS the full source of secondary particles consists of
smaller sources generated in individual $N$-$N$ collisions. Let
$s_{c}(x,p;x_0)$ be the emission function describing such a small
source where $x_0$ stands for the point where the $N$-$N$ collision
occurs. Translation invariance gives $s_{c}(x,p;x_0) =
s_{c}(x-x_0,p)$, and the total emission function can be written as 
 \begin{equation}
 \label{2.1}
   S(x,p) = \int  d^4x_0 \, \sigma(x_0)\, s_{c}(x-x_0,p) \, ,
 \end{equation}
where $\sigma(x_0)$ is the distribution of collision points resulting
from the collision geometry. Its normalization $\int d^4x_0\,
\sigma(x_0)$ is fixed by the number of $N$-$N$ collisions.  

We are now interested in the $q_l$-dependence of the correlator. 
Inserting (\ref{2.1}) into (\ref{1a}), setting $q_\perp = 0$, and 
boosting into the LCMS ($\beta_l=0=K_l$) we find 
 \begin{eqnarray}
    &&C(q_l,K_\perp)\Bigr|_{q_\perp=K_l=0} - 1 
 \nonumber \\
    && = \frac{\left | \int  d^4x \, d^4x_0 \, e^{-iq_l z}
        \sigma(x_0) \, s_{c}(x-x_0,K)\, \right |^2}{\left | \int
        d^4x \, d^4x_0 \,
        \sigma(x_0) \, s_{c}(x-x_0,K)\, \right |^2}
 \label{2.2}\\
    && = \left | \frac{\int d^4x \, e^{-iq_l z} s_c(x,K)}{\int d^4x \,
        s_c(x,K)} \right |^2 \,
        \left | \frac{\int d^4x \, e^{-iq_l z} \sigma(x)}{\int d^4x \,
        \sigma(x)} \right |^2 \, 
 \label{2.3}
 \end{eqnarray}
where $z=\tau\,\sinh\eta$. This expression is typical for LEXUS type
models where $\sigma(x)$ is determined by collision geometry only. 
However, an additional $K$-dependence of $\sigma$ would not affect the
following argument. The correlator would still factorize, and
$\sigma(x)$ will still be non-zero only in the region where $N$-$N$
collisions occur. 

The correlator (\ref{2.3}) is in fact a product of two correlators 
associated with the sources $s_c$ and $\sigma$, respectively. The
longitudinal radius parameter of the whole source extracted from a
Gaussian fit to (\ref{2.3}) is thus given by 
 \begin{equation}
 \label{2.4}
    R_l^2 = R_{l,s_c}^2 + R_{l,\sigma}^2 \, ,
 \end{equation}
where $R_{l,s_c}$ ($R_{l,\sigma}$) is the longitudinal radius parameter 
associated with the source function $s_c$ ($\sigma$). $R_{l,s_c}$ can
be extracted from correlation measurements in $N$-$N$ collisions. 
Based on data published in \cite{NA22,PCh86,NA22-2} this value was
estimated in \cite{PPZ97} to be $R_{l,s_c} \approx 1.7$ fm. 

When considering different collision systems clearly only 
$R_{l,\sigma}$ changes. It can be calculated using the
model-independent expression \cite{CSH95}
 \begin{equation}
 \label{2.5}
   R_{l,\sigma}^2 = \langle z^2 \rangle_\sigma 
   - \langle z\rangle_\sigma^2 \, ,
 \end{equation}
where $\langle f(x) \rangle_\sigma = \int d^4x \, f(x) \, \sigma(x) /
\int d^4x \, \sigma(x)$.

The compilation in Fig.~1 of Ref.~\cite{PPZ97} shows that in
$A$-$B$ collisions $R_{l,\sigma}$ rises approximately linearly with 
$(A-1)^{1/3} + (B-1)^{1/3}$, with a slope which is close to 0.5.
The geometric distributions $\sigma(x)$ for the collision points 
in RWM or LEXUS cannot (via Eq.~(\ref{2.5})) reproduce such a strong rise
\cite{PPZ97}. This can be seen immediately by calculating the r.h.s of
(\ref{2.5}) in an approximation where the two nuclei are replaced 
by cylinders with the same volume and $\langle z^2 \rangle$. One 
finds \cite{PPZ97}
 \begin{equation}
 \label{2.7}
   R_{l,\sigma}^2 = \frac{1}{20} \, \frac1{\gamma^2} \, (R_A^2 + R_B^2),
 \end{equation}
where $\gamma \approx 10$ is the Lorentz contraction factor in the 
$N$-$N$ CMS at CERN/SPS energies. The factor $1/(20 \gamma^2)$
makes $R_{l,\sigma}$ rise by about a factor 6 more slowly than
required by the data; for Pb+Pb collisions, after adding the
contribution $R_{l,s_c}^2 \approx 3$ fm$^2$,
Eqs.~(\ref{2.5},\ref{2.7}) underpredict the data \cite{NA49C} for
$R_l^2$ by a factor 10. Hydrodynamic parametrizations
\cite{CSH95,CNH95} of the source function, on the other hand, can
easily accommodate this rise by adjusting the lifetime $\tau_0$ of the
collision fireball. For Pb+Pb collisions at CERN one finds $\tau_0 
\approx 8$ fm/$c$ \cite{hirschegg97}. This parameter measures the
length of time during which the secondary particles created in the
nuclear collision rescatter among each other. In scenarios with
approximately boost-invariant longitudinal expansion $R_l$ is directly
proportional to this time. Models without rescattering (like RWM and
LEXUS) do not lead to such a collective longitudinal expansion and
thus cannot reproduce $R_l$. 

In summary we conclude that the RWM and LEXUS models, which try to
describe the hadronic momentum spectra in $A$-$B$ collisions in terms
of a linear extrapolation of $pA$ respectively $pp$ spectra, cannot
reproduce the magnitude of the transverse radius parameter $R_\perp$
and its dependence on the transverse momentum nor the dependence of the
longitudinal radius parameter $R_l$ on the size of the collision
system. The first failure is due to the lack of {\em transverse}
expansion and transverse space-momentum correlations in these models.
The second failure is caused by the lack of rescattering which would
result in longer lifetimes and larger values of $R_l$ through
collective {\em longitudinal} expansion. Simple improvements of these
models which preserve their original spirit as a superposition of
$N$-$N$ collisions cannot remedy these failures. Extensive final state
rescattering among the produced hadrons, as implemented in other Monte
Carlo transport models and in hydrodynamical source parametrizations,
is essential to reproduce the correlation data.

This work was supported by DAAD (B.T.), DFG, BMBF, and GSI (U.H.).
We thank Andrei Leonidov for clarifying discussions.



\begin{references}
\bibitem{fn1}
  Note that here the factor ${3\over 4}$ of \protect\cite{LNS97} which
  reflects the impact parameter average has been replaced by the
  correct value ${2\over 3}$.
\end{references}
\end{document}